\documentclass[a4paper,11pt]{article}
\usepackage{pos}

\usepackage{subcaption}

\usepackage{mathtools}
\usepackage{physics}

\usepackage[disable]{todonotes}

\usepackage{verbatim}

\makeatletter
\g@addto@macro\bfseries{\boldmath}
\makeatother

\newcommand{\m}[1]{\mathrm{#1}}

\let\phi\varphi
\let\epsilon\varepsilon

\title{Importance of local tetraquark operators for $T_{cc}(3875)^+$}

\author*[a]{Andres~Stump}
\author[b]{Jeremy~R.~Green}

\affiliation[a]{Institut für Physik, Humboldt-Universität zu Berlin,\\
  Zum Großen Windkanal 2, 12489 Berlin, Germany}

\affiliation[b]{John von Neumann-Institut für Computing NIC,
  Deutsches Elektronen-Synchrotron DESY,\\
Platanenallee 6, 15738 Zeuthen, Germany}

\emailAdd{andres.stump@hu-berlin.de}
\emailAdd{jeremy.green@desy.de}

\abstract{
  The doubly charmed tetraquark $T_{cc}(3875)^+$ observed at LHCb has attracted considerable interest in recent years. To accurately determine its finite-volume spectrum, a variational analysis using a large basis of operators, including bilocal scattering operators, but also local tetraquark operators, should be employed. Using Wilson-clover fermions at the $SU(3)$-flavour-symmetric point, we investigated the importance of local tetraquark operators for the $T_{cc}$ spectrum by adding them to a large basis of bilocal $DD^*$ and $D^*D^*$ scattering operators. We performed this calculation using the distillation framework combined with a position-space sampling method that we recently developed. This method makes local tetraquark operators affordable in distillation. Upon including local tetraquark operators, we observe significant shifts in the estimates of several energy levels. Finally, we show the effect of these shifts on the $DD^*$ scattering phase shifts obtained from a single-channel $s$-wave Lüscher analysis.
}

\FullConference{The 42nd International Symposium on Lattice Field Theory (LATTICE2025)\\
2-8 November 2025\\
Tata Institute of Fundamental Research, Mumbai, India\\}

\begin{document}
\maketitle

\section{Introduction}

In the past two decades, various exotic hadrons have been discovered. A prominent example is the doubly-charmed tetraquark $T_{cc}(3875)^+$ which was observed and studied at LHCb~\cite{LHCb:2021vvq,LHCb:2021auc}. It was discovered in the $D^0D^0\pi^+$ mass spectrum very close to the $D^{*+}D^0$ threshold. This $I(J^P)=0(1^+)$ state has a particularly narrow width, and with a minimal quark content of $cc\bar{u}\bar{d}$, it clearly does not fit into the traditional quark model.

These particle discoveries call for first-principle predictions of their masses, widths and other properties. The rigorous method for doing so is to investigate the poles in the corresponding scattering amplitude. A widely used approach for obtaining the latter is through the use of Lüscher's finite-volume quantization conditions~\cite{Luscher:1986pf,Luscher:1990ux}, which give a relation between the amplitude and the low-lying finite-volume spectrum.

The standard way for computing the finite-volume spectrum is the variational method~\cite{Blossier:2009kd}, which uses a basis of interpolating operators that carry the quantum number of the state of interest. The symmetric correlator matrix of these operators is then utilized to extract the low-lying energy levels by solving a generalized eigenvalue problem (GEVP). To capture the finite-volume spectrum well, a diverse basis of operators should be used. Not doing so can result in inaccurate energy estimates or even in missing energy levels~\cite{Wilson:2015dqa}. As the inner structure of exotic hadrons is generally not known, we want to use operators with different spin, color and spatial structures. For tetraquarks such as the $T_{cc}$, this means we want to use nonlocal meson-meson scattering operators that resemble weakly bound ``molecular'' states, as well as local tetraquark operators for more deeply bound states. Combining these two types of operators in a variational analysis is challenging, since two-point functions of all combinations of operators have to be computed to construct the correlator matrix.

The distillation method~\cite{HadronSpectrum:2009krc,Morningstar:2011ka} is an efficient framework for computing two-point functions of nonlocal scattering operators, and it has been successfully applied for tetraquark states~\cite{Padmanath:2022cvl, Whyte:2024ihh, Prelovsek:2025vbr, Shrimal:2025wbu}. However, local tetraquark operators are computationally expensive within the traditional distillation framework, although such calculations have nonetheless been performed~\cite{Cheung:2017tnt, Ortiz-Pacheco:2023ble,Prelovsek:2025vbr}. Our recently developed position-space sampling method~\cite{Stump:2025owq} makes these operators affordable within distillation.

In this work, we used this position-space sampling method to investigate the importance of local tetraquark operators in combination with bilocal scattering ones for obtaining an accurate finite-volume $T_{cc}$ spectrum. The paper is structured as follows: In Section~\ref{sec:loc_tetraquark_op_in_distill}, we briefly describe how the position-space sampling method is applied for local tetraquark operators. In Section~\ref{sec:T_cc_results}, we show the results from our operator importance analysis for the $T_{cc}$ tetraquark, and present the finite-volume $T_{cc}$ spectra obtained from a purely bilocal operator basis and from a mixed basis consisting of bilocal and local operators. Finally, we present the resulting $s$-wave scattering phase shifts obtained from a single-channel $s$-wave Lüscher analysis before concluding in Section~\ref{sec:Conclusion}.

\section{Position-space sampling for local tetraquark operators in distillation} \label{sec:loc_tetraquark_op_in_distill}

In hadron spectroscopy, quark smearing is crucial for enhancing the overlap of operators with the low-lying states. Within the distillation method~\cite{HadronSpectrum:2009krc,Morningstar:2011ka}, LapH smearing is utilized, which uses the eigenvectors of the spatial, covariant Laplacian $\Delta(t)$ to construct the smearing kernel. More precisely, the smearing kernel $V(t)V(t)^\dag$ is used, where the matrix $V(t)$ contains the lowest $N_v$ eigenvectors of $\Delta(t)$. This makes $V(t)V(t)^\dag$ a projector to the subspace spanned by these eigenvectors. In this so-called LapH subspace, the fermion propagator $S_f = D_f^{-1}$ (with flavour $f$) can be fully computed. This results in the so-called perambulator
\begin{equation}
    \tau_f(t', t) = V(t')^\dag \cdot S_f(t', t) \cdot V(t)
\end{equation}
which is a $4N_v \times 4N_v$ matrix. To maintain a constant smearing radius when changing the volume and lattice spacing, the number of Laplacian eigenvectors has to be scaled proportionally to the physical volume~\cite{Morningstar:2011ka}.

\begin{figure}
  \includegraphics[width=0.5\textwidth, trim={0 -0.4cm 0 0}]{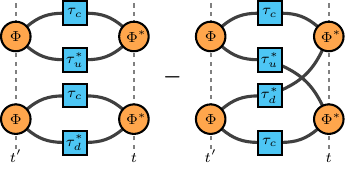}
  \includegraphics[width=0.5\textwidth, trim={-0.6cm 0 -0.6cm 0}]{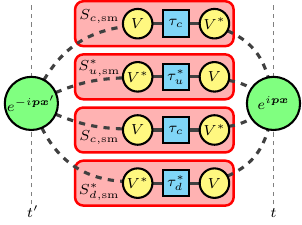}
  \caption{\label{fig:corr_in_distill_and_pos_space_samp}Tensor-network diagrams of two-point functions in distillation of operators relevant for the $T_{cc}$ tetraquark; a bilocal $D^{(*)}D^*$ (left) and a local tetraquark operator (right). For bilocal $D^{(*)}D^*$ operators, the two-point functions are computed by contracting the perambulators $\tau_f$ with the mode doublets $\Phi$. For local tetraquark operators, the smeared fermion propagators $S_{f,\m{sm}}$ are constructed from the perambulators and subsequently contracted with the momentum projections $e^{\pm i\vb*{p}\vb*{x}^{(\prime)}}$ using position-space sampling (indicated by the dashed lines). The spin and color structure is suppressed in these diagrams.}
\end{figure}

Within the traditional distillation framework, two-point functions are computed as tensor contractions of perambulators, gamma matrices and additional tensors which encompass the momentum projection. For bilocal $DD^*$ and $D^*D^*$ operators relevant for the $T_{cc}$, this additional tensor is given by the mode doublets $\Phi(t)$ which form $N_v \times N_v$ matrices. This leads to the tensor-network diagram that is shown in the left panel of Figure~\ref{fig:corr_in_distill_and_pos_space_samp}.
Since it involves only matrix-matrix multiplications in the LapH subspace, the computational cost of the contraction scales as $N_v^3$. This is affordable and, consequently, has been done by various collaborations.

For local tetraquark operators, such as a local $DD^*$ or a diquark-antidiquark operator, the traditional way of computing two-point functions results in a rank-4 tensor that encompasses the momentum projection. As a consequence, the cost of the contraction scales as $N_v^5$, which is expensive, especially in large physical volumes. To circumvent this strong cost scaling, we employ our recently developed position-space sampling method~\cite{Stump:2025owq} to compute these two-point functions within distillation. This leads to the tensor-network diagram shown in the right panel of Figure~\ref{fig:corr_in_distill_and_pos_space_samp}. The basic idea of the position-space sampling method is to avoid the aforementioned rank-4 tensors by first constructing the smeared propagators $S_{f, \m{sm}}$ from the perambulators. They are then contracted with the exponentials $e^{\pm i\vb*{p}\vb*{x}^{(\prime)}}$ arising from the momentum projection, using the appropriate spin and color structure. To make the resulting double sum over position space affordable, we only sum over randomly-displaced sparse grids. The random displacements are sampled for each gauge configuration and source time used. This results in an unbiased stochastic estimator for the two-point function that is different from the one in the traditional distillation method. As for the bilocal meson-meson operators, the computational cost of these two-point functions scales as $N_v^3$. However, the computational cost also crucially depends on the point separation $N_\m{sep}$ in the sparse grids, i.e. the distance between the points in lattice units. As a function of $N_\m{sep}$ the cost scaling is $N_\m{sep}^{-6}$. In~\cite{Stump:2025owq} we determined the ideal point separations for the gauge ensembles used in this work, i.e. the value for $N_\m{sep}$ for which the computational cost is smallest, while still having a variance that is dominated by the Monte Carlo error.

\section{Importance of local tetraquark operators for $T_{cc}$} \label{sec:T_cc_results}

In this section, we investigate how important local tetraquark operators are for an analysis of the $T_{cc}$. First, we examine how the estimated finite-volume spectrum changes when local tetraquark operators are added to a purely bilocal operator basis. We then study their effect on the $s$-wave scattering phase shifts obtained from a Lüscher analysis. All errors only include statistical uncertainties computed using the $\Gamma$-method~\cite{WOLFF2004143}.

\subsection{Lattice setup}

For our simulations, we used two CLS~\cite{Bruno:2014jqa} gauge ensembles with $O(a)$-improved Wilson fermions at the $SU(3)$-flavour-symmetric point. We utilized the same fermion action for the valence charm quark, where the mass was tuned such that the $D$ mass matches the physical average of the $D^0$, $D^+$ and $D^+_s$ masses. In this unphysical setup, the pion has a mass of $m_\pi \approx 420$~MeV, which results in a $D^*$ meson that is stable within QCD. This avoids three-particle decays of the $T_{cc}$. We used stout smearing~\cite{Morningstar:2003gk} in the spatial Laplacian. The relevant parameters for the simulations are summarized in Table~\ref{tab:ensembles}.
\begin{table}
  \caption{\label{tab:ensembles}Parameters of the gauge ensembles used for the simulations. The lattice spacing was determined in~\cite{Strassberger:2021tsu}. $N_{\text{cfg}}$ and $N_{\text{src}}$ are the number of gauge field configurations and sources per gauge configuration used.}
  \begin{tabular}{lccccccccc}
    \hline\hline
    Ensemble & $N_s^3 \times N_t$ & $\beta$ & $a$ [fm] & $\kappa_u=\kappa_d=\kappa_s$ & $\kappa_c$ & $N_{\text{cfg}}$ & $N_{\text{src}}$ & $N_v$ & $N_\m{sep}$ \\
    \hline\hline
    B450 & $32^3\times64$  & 3.46 & 0.0749 & 0.136890 & 0.126243 & 1612 & 8 & 32 & 8 \\
    N202 & $48^3\times128$ & 3.55 & 0.0633 & 0.137000 & 0.128423 &  899 & 8 & 68 & 8 \\
    \hline
  \end{tabular}
\end{table}

\subsection{Finite-volume $T_{cc}$ spectrum}

To investigate the importance of local operators for the $T_{cc}$ spectrum, we focused on the rest-frame $T_1^+$ irreducible representation (irrep) of the octahedral group. For that, we used a basis of nine bilocal $D^{(*)}D^*$ operators, which we then extended by three local tetraquark operators, all with isospin $I=0$.

For the bilocal basis, we used $DD^*$ and $D^*D^*$ operators of the form
\begin{align} \label{eq:operator_bilocal_DDstar}
  &\mathcal{O}_{i, \vb*{p}}^{DD^*}(t) =
		\sum_{\vb*{x}_1, \vb*{x}_2 \in \Lambda_3}
		\hspace{-0.2cm} e^{-i\vb*{p}\cdot(\vb*{x}_1 - \vb*{x}_2)}
		(\overline{u} \gamma_5 c)(\vb*{x}_1, t) \; (\overline{d} \gamma_i c)(\vb*{x}_2, t)
    - \{u \leftrightarrow d\}, \\
  \label{eq:operator_bilocal_DstarDstar}
  &\mathcal{O}_{i, \vb*{p}}^{D^*\!D^*}(t) =
		\sum_{\vb*{x}_1, \vb*{x}_2 \in \Lambda_3}
		\hspace{-0.2cm} e^{-i\vb*{p}\cdot(\vb*{x}_1 - \vb*{x}_2)}
		\epsilon_{ijk}(\overline{u} \gamma_j c)(\vb*{x}_1, t) \; (\overline{d} \gamma_k c)(\vb*{x}_2, t).
\end{align}
More precisely, we used one operator for each (degenerate) non-interacting $D^{(*)}D^*$ energy level and organized them in groups that we call $D^{(*)}(\vb*{k}^2)D^*(\vb*{k}^2)\;\{N_\m{ops}\}$ with $\vb*{k} = \frac{L}{2\pi}\vb*{p}$, where $N_\m{ops}$ is the number of operators in this group. For the $DD^*$ operators we used the momentum shells $\vb*{k}^2 = 0, 1, 2$ with $N_\m{ops} = 1, 2, 3$ operators, respectively. For the $D^*D^*$ we used $\vb*{k}^2 = 0, 1$ with $N_\m{ops} = 1, 2$, respectively.

To this basis, we added the following three local tetraquark operators
\begin{align} \label{eq:operator_local_DDstar}
  &T_i^{DD^*}(t) = \sum_{\vb*{x} \in \Lambda_3}
      (\overline{u} \gamma_5 c \; \overline{d} \gamma_i c)(\vb*{x}, t)
      - \{u \leftrightarrow d\}, \\
  \label{eq:operator_local_DstarDstar}
  &T_i^{D^*D^*}(t) = \sum_{\vb*{x} \in \Lambda_3}
      \epsilon_{ijk}(\overline{u} \gamma_j c \; \overline{d} \gamma_k c)(\vb*{x}, t), \\
  \label{eq:operator_local_diq}
  &T_i^\m{diq}(t) = 
		\sum_{\vb*{x} \in \Lambda_3}
		(\epsilon_{abc} \, c_b^T \, C\gamma_i \, c_c \;
		\epsilon_{ade} \,\overline{u}_d \, C\gamma_5 \, \overline{d}_e^T)(\vb*{x}, t).
\end{align}
The first two are the local $DD^*$ and the local $D^*D^*$ operators, respectively, and the last one is the diquark-antidiquark operator in the $(\vb*{\overline{3}}_c \otimes \vb*{3}_c)_{\vb*{1}_c}$ color irrep. We call the group of local tetraquark operators $T\;\{3\}$.

To extract the low-lying energy levels $E_n$, we used the variational method outlined in~\cite{Blossier:2009kd}. When solving the GEVP, we set $t_0/a = \lceil t/(2a)\rceil$ where $\lceil\cdot\rceil$ is the ceiling function. That way, the condition $t_0 \geq t/2$ holds which ensures a suppression of the excited states with $O\big(e^{-(E_{N_\m{ops}}-E_n)t}\big)$ in the effective energies $E_{\m{eff},n}(t)$. Here, $N_\m{ops}$ is the number of operators used in the GEVP. For the $T_{cc}$, the finite-volume energy levels are strongly correlated with the non-interacting ones. Therefore we extracted the $E_n$ by performing plateau fits to the effective energy difference ${\Delta E_{\m{eff},n}(t) = E_{\m{eff},n}(t) - m^D_\m{eff}(t) - m^{D^*}_\m{eff}(t)}$, where $m^D_\m{eff}$ and $m^{D^*}_\m{eff}$ are the effective masses of the $D$ and the $D^*$ respectively. This results in a significant error cancellation in the fit result. To avoid fake plateaus that can occur in the effective energy difference, we only fitted in regions where both the two- and the one-particle effective energies have reached a plateau.

\begin{figure*}
  \includegraphics[width=0.5\textwidth]{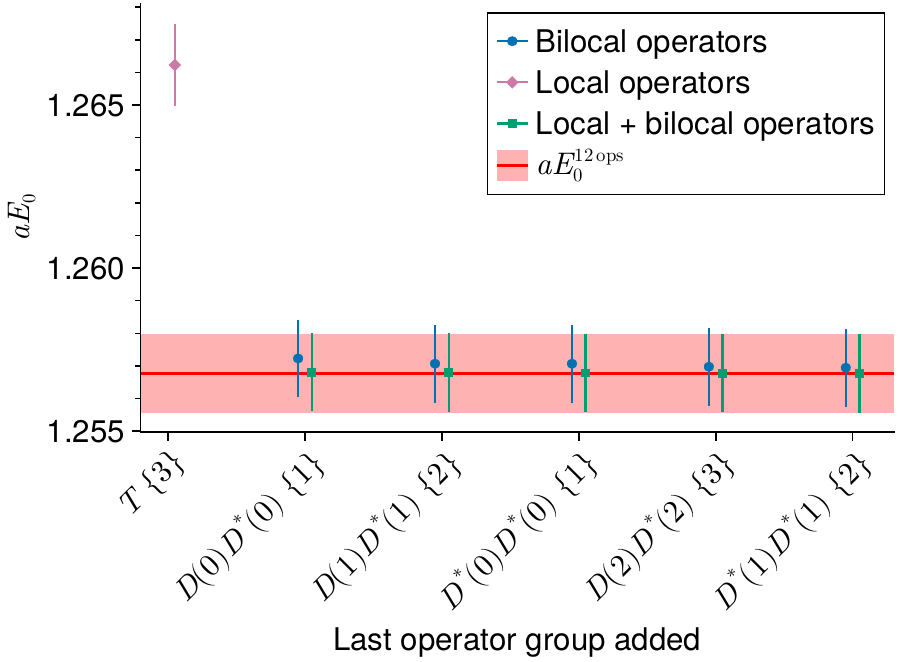}%
  \includegraphics[width=0.5\textwidth]{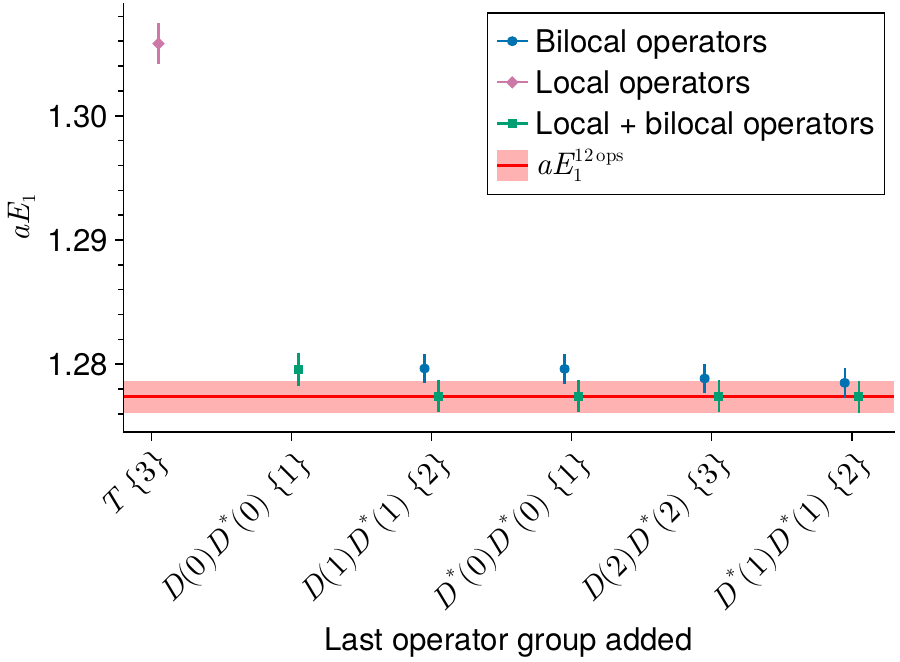}
  
  \includegraphics[width=0.5\textwidth]{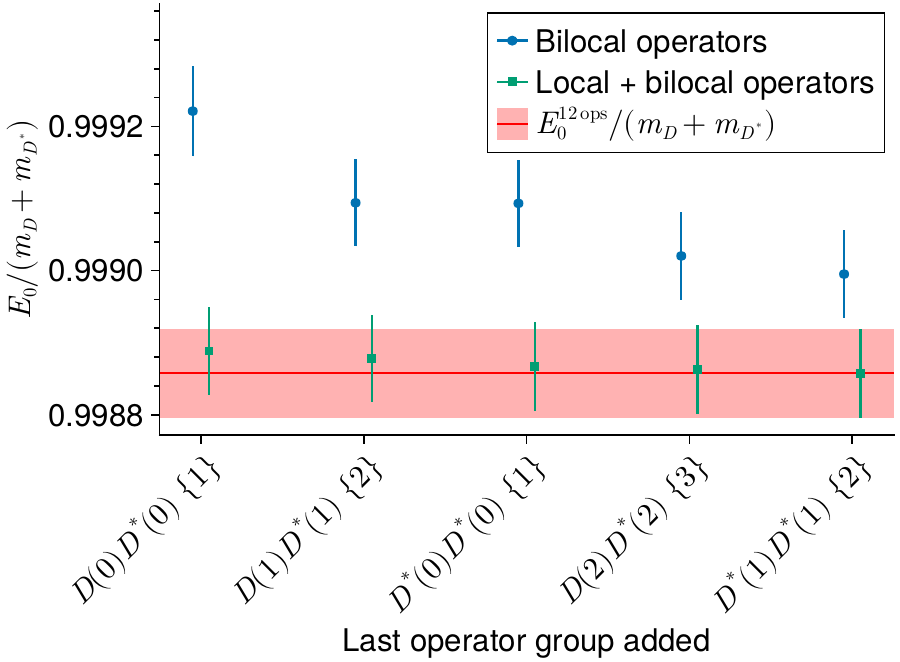}%
  \includegraphics[width=0.5\textwidth]{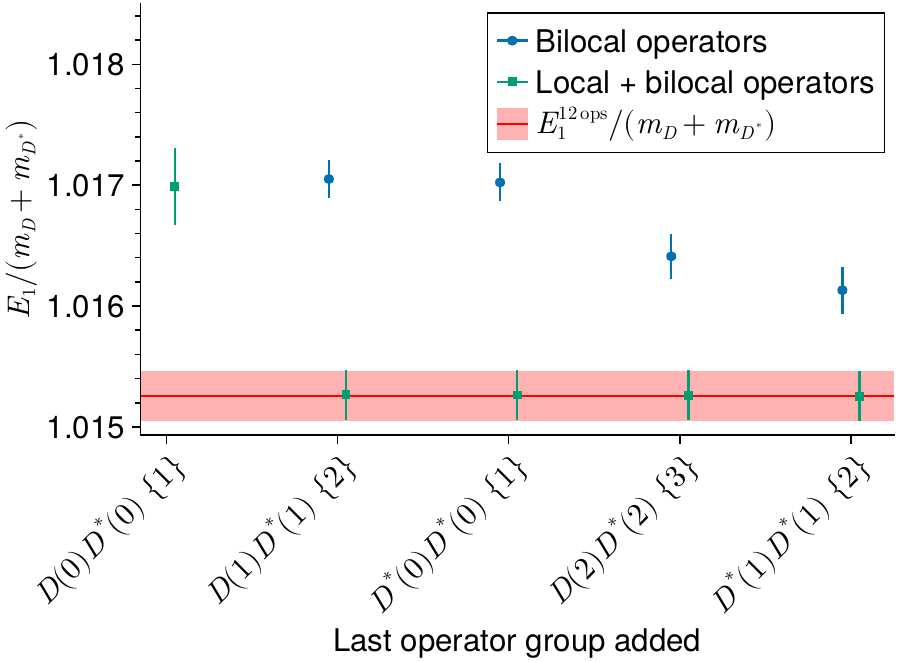}
  \caption{\label{fig:operator_importance}Finite-volume ground state (top left) and first excited state (top right) energy estimates from using three different operator bases: only local operators, only bilocal operators, and local and bilocal operators. The x-axis denotes the last operator group added to the basis for computing $E_n$; groups to the left were already present (except $T\;\{3\}$ when only bilocal operators are used). The red bands show the results from using all 12 operators. The lower panels show the same results, but the energy estimates are normalized with the threshold energy using correlated ratios. We removed the $T\;\{3\}$ energy estimates from the lower plots due to their large values compared to the others. All energies were computed on the N202 gauge ensemble.}
\end{figure*}

To investigate the effect of the different operators systematically, we enlarged the operator basis one operator group at a time, and in each step, we extracted the lowest two energy levels (or just the ground state if there was only one operator). We extended the operator basis in the following order: First the three local operators (if included) and then the bilocal operator groups ordered by increasing non-interacting $D^{(*)}D^*$ energy. The results obtained on N202 are displayed in the upper two panels in Figure~\ref{fig:operator_importance}.
For the ground state, we see agreement between the bilocal and the mixed basis, also when only using one bilocal operator. For the first excited state, we see a fast convergence of the energy estimates when using both local and bilocal operators. However, for the bilocal basis, the convergence is slower, and a 1~$\sigma$ discrepancy remains when using all bilocal operators. We also see that for both the ground and the first excited state, the three local operators themselves do not capture the energy levels well.

This seems to indicate that local operators are not particularly relevant for a $T_{cc}$ analysis. However, in a Lüscher analysis, the correlation between the interacting and the non-interacting two-particle energy levels is relevant, as it can lead to a substantial error reduction in the result. To take this into account, we analyzed the correlated ratios between the energy levels and the threshold energy $m_D + m_{D^*}$. These energy ratios are displayed in the lower panels of Figure~\ref{fig:operator_importance}. We see a significant error cancellation in the ratios compared to the energy levels themselves. This allows us to resolve the difference between the estimates from the bilocal and the mixed basis more clearly. Including local tetraquark operators leads to a rapid convergence of the energy levels upon enlarging the basis. This is the case for both the ground and the first excited state. In contrast, when only using bilocal operators, the estimates show a step-like decrease when extending the basis. When using all bilocal operators, there remains approximately a 2~$\sigma$ and a 3~$\sigma$ difference for the ground and the first excited state, respectively, compared to the mixed basis. It is possible that adding more bilocal $D^{(*)}D^*$ operators on higher momentum shells would decrease the resulting energies from the bilocal basis further. However, these operators become increasingly expensive with higher momenta due to the large number of momentum combinations.

\begin{figure*}
  \includegraphics[width=0.5\textwidth]{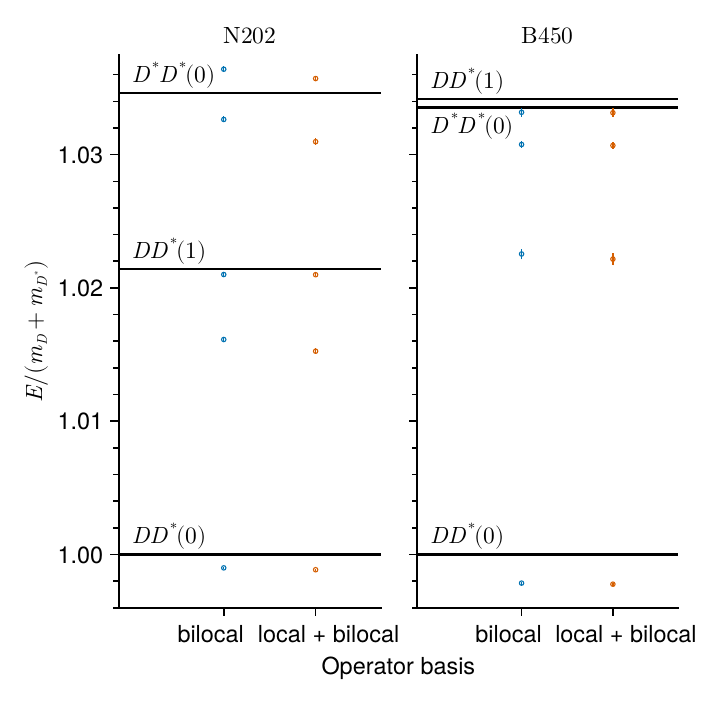}
  \includegraphics[width=0.5\textwidth, trim={0 -1.5cm 0 0}]{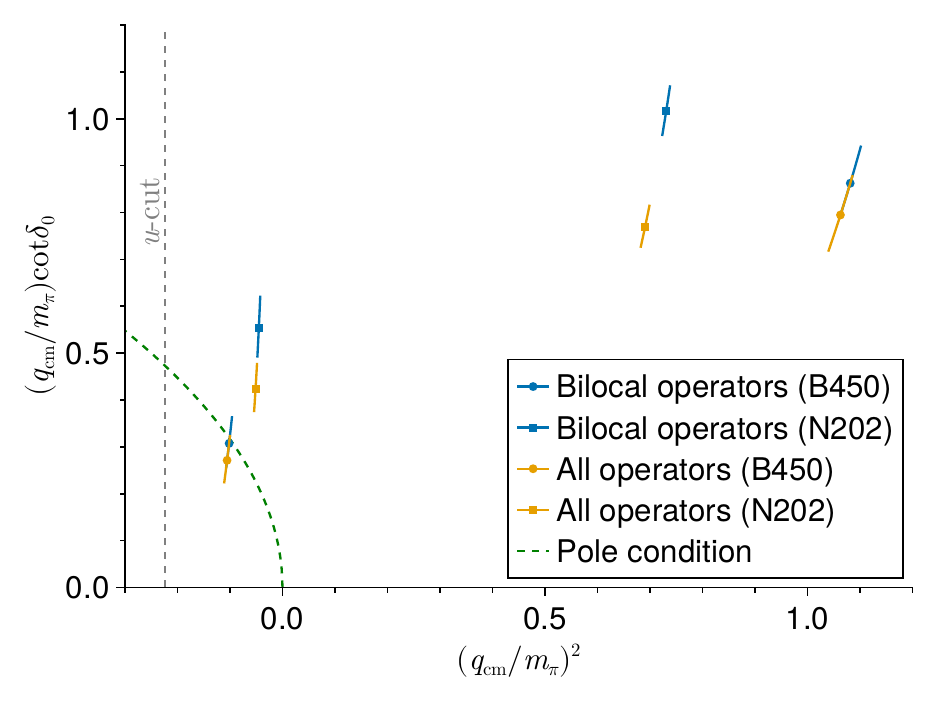}
  \caption{\label{fig:spectrum_and_s-wave_analysis}Left panel: Estimate of the finite-volume $cc\overline{u}\overline{d}$, isospin-0 spectrum in the $T_1^+$ irrep for the operator basis consisting only of bilocal scattering operators and for the full operator basis including also local tetraquark operators. The energy estimates are normalized with the threshold energy using correlated ratios. The spectrum is shown for the N202 and the B450 gauge ensembles. The horizontal lines denote the non-interacting $D^{(*)}D^*$ energies with the integer back-to-back momentum squared in the parentheses. \\
  Right panel: Estimate of the $s$-wave $DD^*$ scattering phase shifts shown as $q_\m{cm}\cot{\delta_0(q_\m{cm})}$ extracted from the finite-volume spectra for both operator bases. The c.m. scattering momenta $q_\m{cm}$ are normalized with the pion mass $m_\pi$. The green dashed line displays the pole condition $q_\m{cm}\cot{\delta_0(q_\m{cm})} = \pm \sqrt{-q_\m{cm}^2}$ and the vertical gray dashed line shows the start of the $u$-channel cut.}
\end{figure*}

Using the basis of all bilocal operators and that of all bilocal and local operators, we computed the low-lying finite-volume spectrum on both the N202 and the B450 ensembles. The results are displayed in the left panel of Figure~\ref{fig:spectrum_and_s-wave_analysis}.
For the N202, we again see the shifts in the lowest two levels that we presented already. Besides them, there are also significant shifts in the third and fourth excited states. For the B450, the shifts in the energy estimates are smaller. This might be caused by the smaller volume, which results in a spectrum that is less dense.

We can conclude that both the bilocal and the full operator basis qualitatively produce the same finite-volume spectrum, i.e. there is no level missed. However, depending on the gauge ensemble, including local tetraquark operators can result in significant shifts in the estimates of several energy levels. Consequently, not including these operators can lead to a considerable systematic error.

\subsection{$s$-wave Lüscher analysis}

Finally, we performed a single-channel $s$-wave Lüscher analysis of the obtained spectra to determine the $s$-wave $DD^*$ scattering phase shifts $\delta_0$. In this case, there is a one-to-one correspondence between the center-of-mass (c.m.) scattering momentum $q_\m{cm}$ and $\cot{\delta_0(q_\m{cm})}$ via the generalized zeta function~\cite{Briceno:2014oea}. In the rest frame, this relation is given by
\begin{equation}
  q_\m{cm}\cot{\delta_0(q_\m{cm})} = \frac{2}{\sqrt{\pi}L} Z_{00}\left(1, \frac{q_\m{cm}L}{2\pi}\right),
\end{equation}
where $q_\m{cm}$ is given in terms of the (c.m.) energy $E$ via $E = \sqrt{q_\m{cm}^2 + m_D^2} + \sqrt{q_\m{cm}^2 + m_{D^*}^2}$.

On both ensembles, we measured a deviation from the continuum dispersion relation for the $D$ and the $D^*$ meson. To correct for this, we used the modified dispersion relation $\lambda \vb*{p}^2 + m^2$, with a parameter $\lambda \leq 1$. This modification changes the definition of $q_\m{cm}$ and has already been used in~\cite{Shi:2025ogt}. We obtained the parameters $\lambda_D$ and $\lambda_{D^*}$ for the two mesons from constant fits to $\big(E_{D^{(*)}}(\vb*{p})^2 - m_{D^{(*)}}^2\big)/\vb*{p}^2$ where $E_D(\vb*{p})$ and $E_{D^*}(\vb*{p})$ are the $D$ and $D^*$ energies for momentum $\vb*{p}$. The resulting values are $\lambda_D = 0.968(4)$ and $\lambda_{D^*} = 0.965(6)$ on N202, and $\lambda_D = 0.928(7)$ and $\lambda_{D^*} = 0.91(1)$ on B450.

The results for the $DD^*$ scattering phase shifts are displayed in the right panel of Figure~\ref{fig:spectrum_and_s-wave_analysis}. As expected, the shifts in the energy-level estimates upon including local tetraquark operators also appear in the phase shifts. When using the full operator basis, the phase shifts are in better agreement between N202 and B450. The results without local operators would have suggested a large discretization effect, and this is reduced now. Our results indicate that the $T_{cc}$ is a virtual bound state at this pion mass, as the phase shifts intersect with positive $\sqrt{-q_\m{cm}^2}$. However, the nearby left-hand cut ($u$-cut) may distort this~\cite{Du:2023hlu, Prelovsek:2025vbr, Hansen:2024ffk}.
We plan to investigate this in the future.

\section{Conclusion} \label{sec:Conclusion}

We have used a position-space sampling method within distillation to investigate the importance of local tetraquark operators for the finite-volume $T_{cc}$ spectrum. To this end, we analyzed how the energy-level estimates change when enlarging a basis of bilocal meson-meson operators and a basis consisting of both local tetraquark and bilocal meson-meson operators. We found qualitative agreement between the energy estimates from the two bases. However, we observed a fast convergence when using the mixed operator basis, whereas when using the purely bilocal basis, the estimates converge more slowly. Depending on the gauge ensemble, this results in significant shifts in the estimate of the finite-volume spectrum upon including local tetraquark operators. Finally, we performed a single-channel $s$-wave Lüscher analysis for the spectra obtained from the two operator bases to compare the resulting $DD^*$ phase shifts. We conclude that not including local tetraquark operators in a $T_{cc}$ analysis can lead to significant systematic errors.

\begin{acknowledgments}
AS is grateful to God for support and useful input throughout this project. We thank Renwick J. Hudspith for code development, and M. Padmanath and Fernando Romero-López for crosschecks.  We are also grateful to our colleagues within the CLS initiative for sharing ensembles. Calculations for this project used resources on the supercomputers JURECA~\cite{krause2018jureca} and JU\-WELS~\cite{krause2019juwels} at Jülich Supercomputing Centre (JSC). The raw distillation data were computed using QDP++~\cite{Edwards:2004sx}, PRIMME~\cite{PRIMME}, and the deflated SAP+GCR solver from openQCD~\cite{openQCD}. Contractions were performed using TensorOperations.jl~\cite{TensorOperations.jl} and ITensors.jl~\cite{ITensor}, the Monte Carlo analysis was done using ADerrors.jl~\cite{ADerrors.jl}, and the plots were prepared with Makie.jl~\cite{Danisch2021}. AS's research is funded by the Deutsche Forschungsgemeinschaft (DFG, German Research Foundation) - Projektnummer 417533893/GRK2575 ``Rethinking Quantum Field Theory''.
\end{acknowledgments}

\bibliographystyle{JHEP}
\bibliography{References}

\end{document}